\documentclass[12pt]{article}

\usepackage{times}
\usepackage{helvet}
\usepackage{courier}
\usepackage[hyphens]{url}
\usepackage{graphicx}
\usepackage{subcaption}
\usepackage{amsmath}
\usepackage{wrapfig}
\usepackage{graphicx}
\usepackage{color}
\usepackage{url}
\usepackage{float}
\usepackage{todonotes}
\usepackage{comment}
\usepackage{listings}
\usepackage{color}

\definecolor{mygreen}{rgb}{0,0.6,0}
\definecolor{mygray}{rgb}{0.5,0.5,0.5}
\definecolor{mymauve}{rgb}{0.58,0,0.82}

\usepackage{pifont}

\usepackage{biblatex} %
\usepackage[top=2cm,bottom=1.5cm,left=2.5cm,right=2cm]{geometry}

\usepackage{graphicx}
\usepackage{txfonts}
\usepackage{hyperref}
\usepackage{lineno}
\usepackage[figuresright]{rotating}

\usepackage[utf8]{inputenc}

\definecolor{GreenYellow}{RGB}{210,255,92}

\newcommand\commentout[1]{}

\graphicspath{ {images/} }
\usepackage{listings}
\usepackage{xcolor}

\definecolor{frameColor}{RGB}{204,204,204}

\usepackage{xcolor}
\usepackage{float}

\definecolor{ColorNodeName}{RGB}{140, 56, 99}
\definecolor{ColorNodeInvariant}{RGB}{167, 66, 168}
\definecolor{ColorEdgeGuard}{RGB}{66, 168, 72}
\definecolor{ColorEdgeProbability}{RGB}{168, 122, 66}
\definecolor{ColorEdgeUpdate}{RGB}{66, 66, 168}
\definecolor{ColorEdgeSynchronization}{RGB}{66, 160, 168}

\usepackage{fontenc}

\title{Testing Reactive Systems Using Behavioural Programming, a Model Centric Approach}
\author{Yeshayahu Weiss$^1$\thanks{weissye@post.bgu.ac.il}}

\date{
	$^1$Ben-Gurion University \\%
}

\begin{document}
\maketitle

\begin{abstract}
Testing is a significant aspect of software development. As systems become complex and their use becomes critical to the security and the function of society, the need for testing methodologies that ensure reliability and detect faults as early as possible becomes critical. In academia and in industry, different methods are being developed for improving the testing process. The most promising approach is the model-based approach where a model is developed that defines how the system is expected to behave and how it is meant to react. The tests are derived from the model and an analysis of the test results is conducted based on it.

In the proposed doctoral research, we will investigate the prospects of using the Behavioural Programming (BP) modeling approach as an enabler for a model-based testing approach that we will develop. We will develop a natural language (textual and/or graphical) for representing the requirements. The model users will create with our language will be fed to algorithms that we will develop. This includes algorithms for the automatic creation of minimal sets of test cases that cover all of the system's requirements, algorithms for analyzing the results of the tests, and other tools that support the testing process.

The focus of our methodology will be to find faults caused by the interaction between different requirements in ways that are difficult for the testers to detect. Specifically, we will focus our attention to concurrency issues such as deadlocks and logical race condition.
We will use a variety of methods that are made possible by BP, such as non-deterministic execution of scenarios and use of in-code model-checking for building test scenarios and for finding minimal coverage of the test scenarios for the system requirements using Combinatorial Test Design (CTD) methodologies. We will develop a proof-of-concept tool kit which will allow us to demonstrate and evaluate the above mentioned capabilities. We will compare the performance of our tools with the performance of manual testers and of other model-based tools using comparison criteria that we will define and develop.

This proposal also includes a description of some preliminary work. As elaborated in the proposal, we validated that a BP based modelling language for testing allows for effective generation, execution, and analysis of tests for two small systems with which we have experimented (a model of a telephony system and the Moodle education platform), each tested in a different way. In addition, as part of this research proposal, we checked the matter of covering all requirements using test scenarios efficiently and minimally using CTD methodologies.
\end{abstract}


\section{RESEARCH OBJECTIVES}
Our thesis is that a comprehensive system testing methodology based on behavioural programming (BP), supported by the right algorithms, can increase the reliability of reactive systems as well as reduce the effort required in system testing, thereby reducing the total development effort. 

We will prove that it is possible to develop a modelling language that allows stakeholders to more effectively describe requirements and enable algorithms that exhaustively test systems. We will prove that our approach allows discovery of issues and faults that are usually missed by conventional methods. Our end goal is to shift the testing methodologies left towards requirements. With the modelling techniques and analysis algorithms that we will develop, testers will focus on requirements and the tests will automatically be generated from their models. We will focus on catching bugs that are triggered by unusual ordering of events that programmers may not consider, e.g, when a student changes her last name in the middle of the semester after getting married. 

\underline{Motivation} Scientists and engineers have developed testing tools and methodologies for many years now, but systems with critical bugs are still going to market. As software systems serve in critical, life-threatening tasks, such as autonomous cars, medical devices, and nuclear factories, these bugs must be identified as soon as possible. Our initial review on issues and reports in bug tracking systems like JIRA and GITHUB revealed that many bugs not caught by QA that were reported by end-users are due to unusual sequences of events (e.g., logical race conditions). 

Current testing methodologies and tools focus mostly on test automation~\cite{Mahajan2016}. Testers plot usage stories and the tools give them the ability to execute these stories manually or automatically using recording facilities and scripting languages. This is problematic in our mind because human coverage is limited and cannot cope with the growing complexity of software systems. This problem can potentially be solved by model based testing (MBT)~\cite{Dias-Neto2009} in which the tests are generated from a model, but current MBT methods are too complex for engineers and are not focused on requirements~\cite{Gurbuz2018}. In our work, we will propose languages and tools for developing, maintaining, and for executing models that are aligned with the requirements of the system.

\begin{figure}[h]
\centering
\includegraphics[scale=0.6]{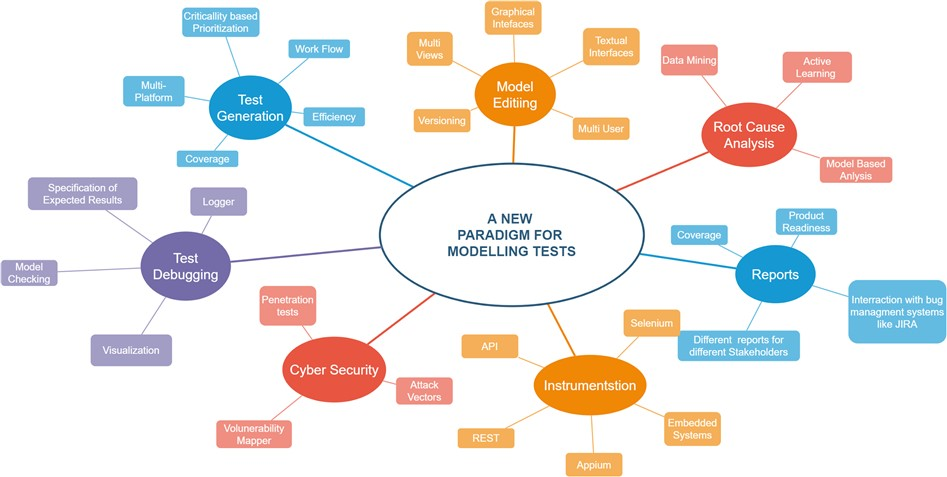}
\caption{An illustration showing how our research will focus on the development of a new modelling technique and will study how it affects the testing workflow.}
\label{fig:brain}
\end{figure}

Towards  improvement of quality and efficiency of testing we propose to improve each of the following ingredients: 1) Modelling, whereby testers design models of all required tests rather than only specifying a set of test scenarios; 2) Automatic generation whereby an engine generates effective test suites from the model; 3) Analysis whereby  absence of contradictions and other properties of the model are validated; 4) Quality measurements whereby algorithms analyse the model and the results of the tests, and generate reports that indicate, e.g., how ready the product is for production; and 5) Prioritization whereby language features and algorithms make decisions about test scheduling, e.g., which tests run nightly?, before product release? after a specific change of the software? All of these issues will be studied with reference to the modelling techniques that we will develop, as shown in Figure~\ref{fig:brain}. We are aware that a full study of all of these topics is beyond the scope of a single Ph.D. research study. Our focus is the modelling technique. We listed the other topics because our research will focus on studying how our new modelling methodologies and techniques reflect on the other issues and because we plan to adjust the modelling techniques in order to better support all software development phases and aspects. 

\subsection{Behavioural Programing methods}
\begin{figure}[h]
\centering
\includegraphics[scale=0.6]{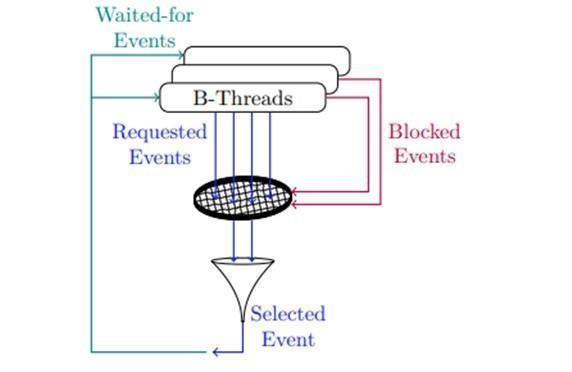}
\caption{BP “life cycle" diagram.}
\label{fig:bplc}
\end{figure}

We will base our research on the Behavioural Programming (BP) modelling approach, whose principles and operations are described in~\cite{Harel2010}. In short, a BP model consists of components called b-threads, each representing an individual aspect of behaviour or scenario that corresponds to a unique paragraph in  the requirement document of a system (if such exists). The b-threads use a special application programming interface (API) that allows them to synchronize with each other in a way that induces a cohesive system behaviour. Specifically, whenever a b-thread reaches a synchronization point (called b-sync), it posts a synchronization statement and waits for all other b-threads to reach their next synchronization points. At synchronization points, b-threads specify three sets of events: Requested events that the thread proposes to be considered for triggering, and asks to be notified when any of them occurs; Watched or waited-for events that the thread does not request but asks to be notified when any of them is triggered; and Blocked events that the thread currently forbids. When all b-threads are at a synchronization point, a central mechanism uses the specified sets to determine the next triggered event, as follows. It selects one event from the set of requested and not blocked events. This selection can be random, priority based or based on an elaborate selection mechanism using, e.g., AI. The selected event is triggered by resuming all the b-threads that either requested it or waited for it. The resumed b-threads proceed with their execution to their next synchronization point, while the other b-threads remain at their last synchronization point, oblivious to the triggered event, until an event they requested or are waiting for is selected. When all b-threads are again at a synchronization point, the process repeats. The BP execution cycle is shown in Figure~\ref{fig:bplc}.

We propose to use BP as an enabler for an approachable model-based testing methodology. We will show how the inherent features and characteristics of BP allow for effective modelling, execution, and analysis of tests. To this end, we will use Context Oriented BP (COBP)~\cite{Elyasaf2021} which combines BP with context idioms that explicitly specify when b-threads are activated and what information they need for their operation. COBP connects the behavioural model with a data model that represents the context. Specifically, COBP consists of an intuitive connection between the data and the behavioural models via update and select queries. For example, a b-thread that represents a requirement concerning a quiz (thing, for example, of a system that tests the Moodle education platform) has access to the properties of the quiz via a designated query protocol. In addition to adding a data model, COBP also allows effective analysis of the testing model by explicitly mapping which b-threads run in a given context. This allows us to avoid analysis of theoretical paths that may not be relevant in certain contexts. COBP maintains dynamic context data that may change at each synchronization point. The b-threads and advanced event selection mechanisms can use this data to direct their internal logic. 

Another BP tool upon which we will base our research is the model-checking mechanism built in current BP tools. For example the BPjs library that implements BP in JavaScript can “dry run” BP models and expand all possibilities of the test scenarios using, e.g., Depth First Search (DFS) of the execution graph. We will use and expand this capability for implementing the model analysis algorithms described above.
   
\section{Planned contributions:}
All of our contributions will revolve around the development of new methodologies for behavioural model-based testing of reactive systems. The following list details some of the aspects on which
we will focus:
\begin{enumerate}
\item \underline{Accessible executable formal modelling languages} for requirement and specification definition. These languages will facilitate the translation of requirement specifications, often maintained by different stakeholders, to test cases. We will develop languages that are: readable – we will make sure that all stakeholders can read the specification so it can be used for feedback and discussion; the language will be rich enough that it can express all tests that will be needed for reactive system testing; and based on BP with its advantages. The models formed by our language will allow a natural expression of system requirements or system specifications on one hand and will cater to automatic generation of tests on the other hand. Our Domain Specific Languages (DSL) will include diagrammatic and textual dialects to ease inter communication between all the teams and individuals involved in the testing process.
\item \underline{Automatic generation of tests}: defining test scenarios based on a behavioural model of the requirements of a system and using algorithms that we will develop based on BP technologies will automatically generate quality test cases. The algorithms that we will develop will be based on expanding as a graph where each b-sync point is a node (state) and each requested and non-blocked event is an edge that represents a transition from one state to another. A path in this graph can be translated to a test case, thus the graph as a whole represents all possible test cases. A subgraph can describe test cases which test a sub-system or a module, or can describe the test cases required for integration test cases between two or more modules in the system. 
\item \underline{A methodology for focusing and coverage}: specification idioms and algorithms for prioritizing tests in order, e.g., to maximize the probability of catching a certain type of bug at the focus of a certain testing effort. This can be done, for example, by applying t-way methods that make use of the fact that most bugs are caused by the interaction of a small number of parameters~\cite{Kuhn2004}. More generally, we will develop mechanisms for managing the choice of which tests to run. This depends on purpose and resources. For example, people run different tests in full system testing, in regression testing, in daily or nightly testing, and in smoke tests. The choice of tests can also depend on the scope: are we focusing on a specific sub-system or on a certain section of the requirements document? We will develop algorithms to choose tests in a way that increases the coverage criteria needed for each type of test. We will also develop methods for measuring different kinds of coverage criteria and for reporting them to users in a useful manner.
\item \underline{Modularity}: we will develop approaches that allow the addition of new requirements that are consistent with already existing ones without touching the parts of the models that already reflect the existing requirements, simply by adding new b-threads. The new test cases will be automatically woven with existing ones to generate tests that examine new possible interactions. This type of modularity contributes to the proposed methodology the ability to advance the development of the system step-by-step, and at each step add the system requirements with the corresponding test cases. These test cases will cover the new requirements and the interaction of older requirements with them. 
\item \underline{Reports and other debugging tools}: we will develop algorithms and tools for analysing test results. For this part of the research we will develop tools and algorithms in domains such as: logging, visualization, playback, summarization, state-based coverage measurement, etc. We will develop methods for analysis of the test cases generated to allow debugging and validation and for checking if the test cases really test what they intend to test based on the system requirements or specification. We will also develop tools that export the outcome of the tests for external processing, e.g., by Business Intelligence (BI) or Artificial Intelligence (AI) tools.
\item \underline{Algorithms for formal analysis / verification of models based on model checking}, which requires simulating tests offline in an "open-loop" manner. We will develop tools for analysing test plans by examining the graphs of all test cases using model-checking tools and algorithms that check the test cases’ space.
\end{enumerate}

\section{Novelty prospects:} 
We will develop an approachable model-based testing (MBT) technique based on the behavioural programming paradigm. Like existing MBT approaches, we will give our users tools to focus on modelling system requirements. Unlike existing MBT, our models will not be convoluted state machines, they will consist of user stories and scenarios that resemble the test scripts to which test engineers are accustomed (but greatly improve expressive power by explicitly specifying what must, may, and may not happen). 

\section{Background and related work:}
\subsection{Testing background}
Software has a tendency to fail. With the progress of the complexity of software systems, it is becoming progressively hard to guarantee the quality of the software. Since it is generally impossible to verify the nonappearance of bugs in a real program, the main goal of software testing is to find bugs as soon as possible so that they can be fixed with minimal cost~\cite{Felderer2016}. It is therefore important to follow a systematic testing practice with the purpose of increasing product assurance while confirming that the software features are as required. Testing methods can be classified into white-, gray- and black-box testing according to their accessibility. When test cases are designed based on information about how the software has been designed or coded it is called white-box testing~\cite{Atesogullar2019}. When the design of the test cases depends only on the input/output, the behaviour or the functional requirements of the software it is called black-box testing~\cite {Cheng2003}. A mixture of the white- and black-box testing methods, using the advantages of both methods, is called gray-box testing.
The following list of test levels was defined by the International Software Testing Qualifications Board  (ISTQB)~\cite{Sneha2017}: {\bf Unit testing}: testing each hardware or software element; {\bf Integration Testing}: finding faults in the interfaces and in the connections between integrated systems or units; {\bf System Testing}: confirming that the integrated system meets the stated features based on system requirements; {\bf Acceptance Testing}: checking that the system satisfies the acceptance criteria with respect to user needs, requirements, and business processes; {\bf Regression Testing}: testing that the software works as it did before after modifications are done that are suspected to add bugs~\cite{Jamil2017}; {\bf Smoke Testing}: a small group of tests that focuses on the critical level of functionality of the SUT. It runs whenever a new build is created or a new build process runs and verifies that the main functionality is still valid~\cite{Dustin1999}.
Classically, in software testing there is a separation of load testing and stress testing from functional testing~\cite{Krichen2020}. {\bf Load testing}: putting a load on a software system and measuring the system’s response. Such tests are accompanied by tools for monitoring the performance of the system. {\bf Stress testing}: estimating the limits within which the system keeps working when it is exposed to heavy loads or when some of its hardware or software is at risk~\cite{Krichen2020}.  Manual and automated testing can be used together at different stages of software quality verification. The method of automated testing comprises the use of special software tools to execute tests. While there are known weaknesses of automated tests~\cite{Oliinyk2019}, most of the industry is adopting said tests. Still, some programmers think that test automation costs are high relative to their value and that they should be used carefully~\cite{Jamil2017}. Generally, large systems with extensive complexity need test automation with a large return of investments (ROI). See more details about testing background in Appendix A.
\subsection{Testing coverage}
In software testing, coverage known as code coverage or test coverage are important metrics and benchmarks by which to measure test quality. Code coverage is a metric to evaluate how many parts of a program have been tested. It is one form of white-box testing which finds the areas of the program exercised and those that were not exercised by a set of test cases. Different researchers compiled lists of code coverage criteria~\cite{Sneha2017, Kim2021, Horvath2019}. These lists include, for example, Statement coverage; Branch coverage; Function coverage; Loop coverage; Condition coverage; and Finite State Machine coverage. Test coverage is defined as a metric in testing that measures the amount of testing performed by a set of tests related to the system under test (SUT). Test coverage is considered to be black-box testing. Test coverage types are: Features coverage~\cite{Lamancha2010}; Requirements coverage~\cite{Tahat2001}; and Input parameters coverage~\cite{Cheng2003}.

Test case generation based on coverage has advantages and disadvantages. The advantages are as follows: First, reliability seems to increase with test coverage~\cite{Yang1995}; Second, code coverage provides the ability to select a set of tests that significantly improves coverage and prioritize them~\cite{Jones2001}; And third, based on observations in industry, increasing code coverage becomes a motivation for improving tests~\cite{Yang2009}. The disadvantages are: First, the number of test cases that are generated in order to achieve more coverage are growing exponentially, and may be impractical; Second, there is no known underlying theory that predicts how much quality improves with coverage~\cite{Cai2005}; And third, full code coverage (100\%) does not guarantee the absence of defects, especially in the concurrent systems (the main concern in our proposal) when the test cases cover each part of the system, but the system behavioural concurrent does not~\cite{Schwartz2018}. See more details about testing coverage in Appendix A. 
\subsection{Test cases coverage using CTD}
\subsubsection{CTD – background}
Running all possible test cases is impractical in large and complex systems since the total number of possible valid test cases is usually prohibitively large (exponential in the number of requirements). Therefore approaches are needed to generate sets of test cases that are substantially smaller than exhaustive test sets but still “cover” systems' requirements and are effective at detecting faults. Combinatorial Test Design (CTD) is an approach for solving this challenge. The approach is based on modelling a test as a set of parameters, each with a finite set of values, and then sampling the test space by combining possible assignments of values to parameters in a systematic fashion. CTD methods have proven very useful in reducing the number of tests while increasing productivity. The source of this success is as follows. If we assume that all the faults in a system are triggered by a combination of t or fewer combinations of parameter values, then a test suite where each such combination appears in at least one test case is effectively equivalent to an exhaustive test~\cite{Kuhn2004}. CTD methods consist of mathematical computations that yield small test suites that cover all such combinations. Empirical studies about software quality and reliability found that, in reality, most bugs are triggered by very small combinations of parameter values and that CTD improves the effectiveness of bug hunting considerably~\cite{Kuhn2004}. In our research, we will study how these methods can be extended to sequence testing and to coverage criteria that arise in the context of behavioural testing. See more details about CTD background in Appendix A.
\subsubsection{CTD – related works}
Classical CTD is designed for covering parameter values, not different ordering of events. This makes it less effective for testing reactive systems. A variant of CTD called sequence testing addressed this weakness by focusing on t-length sequences of events from a finite set E and then requiring that every such sequence has to occur as a subsequence of at least one test case. Elements of t-length sequences do not have to appear in a sequence in the test case. 

The first variant of sequence testing allowed only one triggering of each event in a test case~\cite{Yu2012}. Later versions allowed more than one triggering in a test case and added support for advanced constraints and other features~\cite{Sheng2018, Duan2019, Bombarda2020}. Current sequence testing methods consist of a two-step method: the first step is to generate a list of all relevant sequences of length t (called 'Target Sequences') and the second step is to generate test cases to cover the list of all target sequences (called 'Test Sequences'). See~\cite{Yu2012}. In the test sequence generation step, the paper starts by using a greedy algorithm that handles constraints between two events, then a transition label system is proposed to represent the SUTs’ requirements and graph path methods are used in order to find the optimal valid test cases. Based on this work, additional work has been done to expand the language of constraints by, e.g., adding the possibility of contiguous values~\cite{Sheng2018} and by allowing more complex relationships between more than two factors~\cite{Duan2019}. Two main problems that remain open are the ability to model the SUTs’ requirements in a way that allows the creation of valid test cases based on t-way testing and that the solution will be effective at run time and size, otherwise the solution will not be applicable in complex systems. A new research study proposes to model the SUT by a finite state machine and to generate the test cases using automata theory~\cite{Bombarda2020}. In each of the papers mentioned above, the researchers present algorithms for generating test cases, they evaluate them and present the results as a number of test cases and their total length. All of these examples demonstrate the fact that the number of generated test cases covering all cases is significantly lower than the overall number of options.

\
In our research, we will continue this line of work by adding new algorithms and coverage criteria that fit the new testing methodology that we are proposing. One specific challenge that arises in our setting is how to take advantage of the modular nature of the model that we are proposing. Specifically, existing methods rely on an analysis of automata and state machines whose sizes grow exponentially with the complexity of the model. In our modelling approach, these state machines are described implicitly as the product of smaller machines. The challenge left for research is to generate effective test suites based on an analysis of the component without an explicit construction of their product whose size is exponentially larger than the sum of their sizes.

\subsection{Model-based testing (MBT)}
\subsubsection{MBT background}
A testing methodology based on a model that defines how the SUT can be interacted with is called Model-Based Testing (MBT)~\cite{Gurbuz2018}. MBT is a black-box testing technique~\cite{Heijden2019}. The general process for MBT is that based on the test requirements and the test plan, a test model is constructed. The test model is used to generate test cases and test oracles. Because of that, there are usually an infinite number of possible tests; usually test selection criteria are adopted to select the proper test cases. The test execution will result in a report that contains the outcome of the execution of the test cases. In the final, these results are analysed and if needed corrective actions are taken. Hereby, for each test that reports a failure, the cause of the failure is determined. The most widely used state-based models in MBT are: finite state machines (FSM)~\cite{Villalobos-Arias2018}, extended finite state machines (EFSM)~\cite{Krichen2020}, UML state machine diagram, timed automata, Markov chain usage models~\cite{Tldx2017}, and labeled transition systems (LTSs)~\cite{Heijden2019}. There is a lack of scientific knowledge regarding these techniques, making it difficult to transfer them to the software industry~\cite{Dias-Neto2010}.

\
MBT limitations and challenges are~\cite {Dias-Neto2009}: {\bf Partial model} - the transition from system specification to a complete model of the system including all interfaces, interactions between the various components and the rest of the relationships, is in many cases incomplete; {\bf Low up-to-datedness model} - The basis on which the model is created (requirements, design, UML, etc.) is in many cases updated during the life of the project, whether due to overload and pressure or for other reasons. The immediate result is that the generated test cases are not actually covered by the system tests; {\bf Skill level} - the skill level required to use MBT approach - knowledge of software modelling notations, test criteria, test metrics, or languages to generate test scripts~\cite{DiasNeto2007}; and {\bf High diversion} - There is a high variance in the characteristics between the software projects and on the other hand there are many academic solutions for MBT~\cite{Dias-Neto2010}. See more details about MBT background in Appendix A.
\subsubsection{MBT related work}
Surveys that were published in recent years~\cite{Villalobos-Arias2018, Bernardino2017, Li2018} presented a homogeneous picture of the existing situation in both academia and industry in the MBT world. There are many tools that present themselves as MBT. The papers identified 70 MBT tools published in 2006-2016, 40 of which are academic tools, 15 are commercial tools, and 15 are open source~\cite{Bernardino2017}. Most tools apply (out of the 5 components that define MBT~\cite{Li2018}) to the creation of the model out of the requirements or specifications and create the test scenarios~\cite{Li2018}; a small portion is also added as a tool for creating the test data and very few tools implement the more complex steps of creating scripts, running the test and a final step of analysing results. These MBT tools model and generate test cases covering functional requirements but not non-functional requirements~\cite{Villalobos-Arias2018}. In addition, the surveys indicate that about 20\% of the MBT products are based on the modelling of the requirements based on UML charts (all types of charts)  and on the modelling of the system with requirements that are expressed as formal or semi-formal modelling~\cite{Bernardino2017}.  In our research, we will propose to develop a new MBT method. Our method will be based on system specifications modelling language. The model that users will create with our language will be fed to algorithms for the automatic creation of minimal sets of test cases that cover all of the system's requirements, automatic execution of the generated test cases, algorithms for analysing the results of the tests, and other tools that support the testing process.
\subsection{Used tool}
\subsubsection{Automatic testing tools}
\underline{Cucumber}~\cite{Cucumber} / \underline{Behat}~\cite{Behat} and \underline{Gherkin}~\cite{Gherkin} - behaviour-driven development (BDD) testing tool. Cucumber is an automatic testing tool that executes software tests in two layers. In the first layer, tests are written in formal language such as Gherkin and the second layer each line in the first layer represented as a function that executes the tests. The second layer supports Java, JavaScript C++ and other languages. Behat is a semi-official BDD automated testing tool like Cucumber for PHP. Cucumber enables automation of functional validation in an easily readable and understandable format (such as plain English) for business analysts, developers, testers, and others. Gherkin is a popular language used by Cucumber to define test cases. Its main objective is to enable users to specify tests in a way that clients can understand them. Gherkin tests are organized into features. Each feature is made up of a collection of scenarios defined by a sequence of steps and following a Given-When-Then (GWT) rule~\cite{RodriguesDaSilva2018}. A simple example is illustrated below. Simple test case example in Gherkin: 

{\em
\hspace{1cm}\color{red}Feature: \color{blue}Login Action

\hspace{1cm}\color{red}Scenario: \color{blue}Successful Login with Valid Credentials

\hspace{2cm}\color{red}Given \color{blue}User is on Home Page
	
\hspace{2cm}\color{red}When \color{blue}User Navigate to LogIn Page 
	
\hspace{2cm}\color{red}And \color{blue}User enters UserName 
	
\hspace{2cm}\color{red}And \color{blue}Password 
	
\hspace{2cm}\color{red}Then \color{blue}Message displayed Login Successfully
}
	
\underline{Selenium} - Selenium~\cite{Selenium} is an object-oriented library for test automation based on browser emulation. It is a suite of tools for automating web application testing across platforms. Selenium runs in several browsers and operating systems and can be used with a variety of programming languages and testing frameworks. The use of Selenium brings many benefits because it allows the use of a common API to control different web browsers. It can be used from the perspective of end users to test applications through the Selenium testing script, and it allows easier detection of browser’s incompatibilities by running tests in different browsers. It simulates the users’ interactive operations with Web applications~\cite{Wang2009}. In our preliminary research we tried to use Cucumber and Gherkin as our language for system requirements modelling language (we describe this in the preliminary result paragraph), but despite the widespread use of these tools in the industry, the vocabulary in this language wasn't rich enough for our purpose.
\subsubsection{SMT solver}
\underline{‘Z3’ SMT solver library}~\cite{Z3Prover} - An SMT solver is a tool for deciding the satisfiability (or dually the validity) of formulas that can handle equality reasoning, arithmetics, fixed-size bit-vectors, arrays, quantifiers, and other useful theories. Given a set of constraints an SMT solver looks for a model that satisfies the constraints or validates that there is no such model. SMT solvers enable applications such as extended static analysis, predicate abstraction, test case generation, and bounded model checking over infinite domains. Z3 is an SMT solver from Microsoft Research. It is targeted at solving problems that arise in software verification and software analysis. Consequently, it supports a variety of theories needed in this domain including the regular expression and string manipulation theories that we have used in our preliminary work. Z3 uses advanced algorithms for quantifier instantiation and theory combination. The first external release of Z3 was in September 2007. Users interact with Z3 using either a textual format or a binary API. Three textual input-formats are supported: The SMT-LIB format, the Simplify format, and a low-level native format in the spirit of the DIMACS format for propositional SAT formulas~\cite{DeMoura2008}. One can also call Z3 procedurally by using either an ANSI C API, an API for the .NET (managed common language runtime) and a Z3 python API called ‘z3py’ (we are using the latter).  

At a high level, the Z3 solver takes as input a logical formula and then tries to decide if the formula is satisfiable. In the process, solvers employ various heuristics that first transform the input formula into a suitable representation and then use search procedures to check for satisfiability. In total, the Z3 SMT solver defines more than 100 such heuristic transformations (called tactics) that can be combined together to define a custom strategy. Although the above sequence of transformations (tactics) works well for some types of input formulas (e.g., in case every variable has a lower and an upper bound), for other formulas a different set of tactics is more suited. In some cases, the suitable set of tactics can be obtained by a small modification of the original tactic while in others a completely different set of tactics needs to be defined~\cite{Balunovic2018}. In the proposed research, we will use Z3 for advanced analysis of the testing models. For example, in a preliminary work, we have used Z3 to compute a set of tests that satisfy certain coverage criteria. For this, we may need to add tactics and theories.
\subsection{Reactive system testing – the challenge}
\subsubsection{Reactive systems testing - background and the challenge}
A reactive system such as an automatic transportation system, a satellite, a drone, or a web application is characterized by the use of on-the-fly of sensors and actuators. Such systems sample the environment at a high rate and produce a rapid response to events. By nature, reactive systems generate a plethora of execution flows that progress simultaneously, as well as concurrency and parallel activities to respond to the complex situation. Traditional testing methods, especially code coverage or code static analysing do not cope well with issues of parallelism and concurrency that cause non-deterministic behaviour and exponential growth of test cases to cover all potential cases. Our research will focus specifically on testing reactive systems and on managing the large space of possible interactions that such systems allow. See more details about reactive system testing background and the challenge in Appendix A.
\subsubsection{Reactive systems testing, related work.}
Researchers have developed techniques that specifically take into account concurrent software features such as non-determinism, synchronization, and communication testing reactive systems. Much of the work in this domain assumes that requirements are specified using formal notation, e.g., Event-B specifications. In~\cite{Vu2017}, for example the authors propose a model-based testing approach where models are Event-B specifications. This approach provides system developers with a template that can generate test scenarios which contain both input values and expected results. Another approach is required when the system has COTS and model-based testing is made more difficult to use directly. In~\cite{Narizzano2020}, for example, the authors propose a methodology that traverses a Büchi automaton that models to the requirements. The traversal starts from the initial state of the automaton and generates a sequence of input values with which the black-box system is fed in order to obtain a corresponding sequence of output values. In~\cite{JianGaoXinYangYuJiangHanLiuWeiliangYing2018}, the authors present a dataset of all cases that can cause race data faults. The dataset contains 985 data race faults, which can be used to evaluate and optimize race detection techniques. The authors also used the dataset to evaluate three race detectors~\cite{JianGaoXinYangYuJiangHanLiuWeiliangYing2018}. Another group of proposed methods deals with safe programming in the sense of interacting with other processes. The approach presented in~\cite{Desai2018}, for example, works towards enabling safe programming of reactive systems. The approach consists of two parts: 1) a programming language for implementing, specifying, and compositionally (assume-guarantee) testing the high-level reactive software; and 2) a runtime verification system to ensure that the assumptions used during design-time hold at runtime. Combining a high-level programming language and its systematic testing with runtime enforcement bridges the gap between software testing that makes assumptions about the low-level controllers and the physical world, and the actual execution of the software on a real platform in the physical world.
\subsection{Behavioural Development}
The need for describing and specifying requirements systems through scenarios and behaviour-driven descriptions has existed for a long time~\cite{Harel2010}. Many techniques, methodologies and tools have been developed throughout the years with varying success. In this work we will use a modelling approach called Behavioural Programming (BP). This an approach that promotes the use of scenarios and anti-scenarios for describing complex behaviours. The approach is based on  Statecharts~\cite{Harel1987} and Life sequence charts (LSC) ~\cite{Damm2001}. See more details about Statecharts, LSC and executable specification in Appendix B.
\subsubsection{BP + COBP}
Describing a system by scenarios and behaviour is a natural way of system description and specification~\cite{Uchitel2003}. BP serves as a link in the transition from behavioural modelling (e.g., LSCs or Statecharts) and behavioural programming in general-purpose programming languages~\cite{Harel2010} such as C++, Java, JavaScript~\cite{Bar-Sinai2018} and more. The BP method is described in Section 1.2.1. BP is an extendable framework. With the basic mechanisms of BP it is possible to define and develop high level structures and design patterns, such as break-upon or interrupt and to extend the language with different modelling idioms. A break-upon pattern, for example, can be added to allow the definition of a structure such as the well-known try-catch idiom used in advanced programming languages, by requesting an event, along with waiting for one or more other events. If the requested event is selected, the process continues the normal activity (try). If the event is not selected, but the process resumes with the event being waited for then the alternative treatment (catch) is caught and treated. Similarly, one can use an interrupt pattern that allows to break the regular flow of the b-thread and to skip to a new flow when some event is triggered. 
BP semantics definition is based on a labelled transition system (LTS)~\cite{Elyasaf2021, Harel2010} where each b-sync point is a state and each event selection is a transition. In general, there may be more than one run of a b-program, depending on the order in which the events are selected from the set of requested and unblocked events. These runs allow designers of systems to separate the specifications of possible behaviours from the process of prioritization and the choice of events. Moreover this allows the b-program to be expressed in the form of an unfolding graph and the program execution between event occurrences is treated as atomic~\cite{Harel2010}.

With BP, specifications of reactive systems are modelled with b-threads that model individual requirements bandeled as a b-program. An obvious limitation of this approach is that requirements sometimes conflict, or are not detailed enough, and composing them automatically without global consideration may yield a composition that produces undesired joint behaviour. The solution to this can come from using the BP a model-checking tool (BPMC). The BPMC tool can verify behavioural programs directly; without translating them into a model-checker-specific language. B-programs can serve both as elements of a final executable system as well as elements of an abstract system model to be subjected to verification~\cite{Harel2011}.

This proposal is based on BPjs framework~\cite{Harel2011}, a platform supporting the growing body of work in behavioural programming under one roof focused on execution and verification of b-programs. BPjs defines a generalized version of BP with well-defined extension points and external interfaces. Thus, BPjs can serve as a common platform for researching and disseminating ideas in BP. BPjs allows b-programs to be embedded in existing software systems by sending data from a host application to the b-program and sending data from the b-program to the host. A super-step based mechanism takes care of embedding the events within the run of the program in a systematic way. BPjs is implemented as a Java library that runs code written in JavaScript. It uses the Mozilla Rhino JavaScript engine to execute regular JavaScript code, and custom code for handling synchronization calls. BPjs framework includes an automatic model-checking tool for verifying the developed software against a formal specification. This tool allows for an exhaustive analysis of the code, producing formal guarantees of quality. 

Context Oriented BP (COBP) combines BP with context idioms that explicitly specify when scenarios are relevant and what information they need. The core idea is to connect the behavioural model with a data model that represents the context, allowing an intuitive connection between the models via update and select queries. Combining BP with context-oriented programming brings the best of the two worlds, solving issues that arise when using each of the approaches separately. COBP is a layer above BP~\cite{Elyasaf2021}. The COBP semantic extends the BP semantic. The life cycle of a context-aware b-program (COBP) is described here. Each context aware b-thread (CBT) is bound to a query on the contextual data. Whenever a new answer exists for a query, new live copies are spawned for the relevant CBTs. The live copies repeatedly execute an internal logic that may depend on the contextual data and then synchronize with each other, by submitting a synchronization statement to a central event arbiter. Once all live copies have submitted their statements, the arbiter selects an event that was requested and was not blocked. The event is also passed to the Effect Function which may update the contextual data, depending on its specification. The (updated) contextual dataset is passed back to the CBTs, along with the selected event. All CBTs that are either requested or waited for this event are resumed, while the rest remain paused until the next cycle~\cite{Elyasaf2021}.

\section{Methods and work plan}
Our work consists of several stages. Each one builds on the results of the previous one. There are three types of stages. 

The first type is pure innovative work - this is the major work in our proposal. This type includes: 
\begin{enumerate}
\item Language (DSL) and / or graphic tool (i.e., Blockly) used by testing development and system engineers (paragraph 2.1) 
\item Using BP concept (such as request, wait for, block, break upon, interrupt) in the processes’ testing and within each action (screen/field) (paragraph 2.2)
\item Control the testing process using break-upon (and context) or block to find out whether either anomaly is a bug or is caused by another process (paragraph 2.6)
\item System Quality measurement and assessment (paragraph 2.5)
\end{enumerate}

The second type is evaluation tools that are required to validate, examine or analyse the results of the first type. This type includes:
\begin{enumerate}
\item How to validate our methodology (paragraph 2.5):
\begin{enumerate}
\item Our methodology vs. conservative methodology using different groups of testers 
\item Find unknown and known bugs in an open source project (i.e., Moodle or openemis) using new methodology and framework
\item Sandbox with planted bugs 
\end{enumerate}
\item Tools for assisting the test cases’ development (i.e. debugger, screenshots, reports) and for validating test cases (paragraph 2.6) 
\item (nice to have) Mathematical analysis - exponential blow-up without blocking 
\end{enumerate}

The third type is implementation framework and tool kits that allow the examination of the applicability and completeness of the results of the previous steps (paragraph 6.3).
	
\section{Preliminary results}
\subsection{Test case coverage – initial results}
\subsubsection{Test case coverage – our suggestion.}
One of the challenges in our research proposal is test case coverage. The research study on this issue and empirical proof of test case coverage methods in Combinatorial Testing Design (CTD)~\cite{Hu2020} are described in paragraph 4.3.1. 

In most of the work on the subject it seems that the barrier of laboratory examples has not yet been breached. The examples that were used by Kuhn et al are good enough and the total impression created on the basis of the results obtained is enough to convince one that the direction of the solution is correct. Even though the examples are minimal, they allow for the possibility of understanding the innovation and the algorithms, but they are not complex enough in relation to composite systems and the models in the real world. Another consequence of this issue is that they bring a small number of examples of very small models. The algorithms that they developed for generating the system model as LTS, automata or other modelling techniques are not applicable when the examples contain more input parameters. In addition, because their samples have relatively few elements and they focused on the methods they developed, they did not pay much attention to the efficiency of the algorithm in terms of complexity and runtime. And finally, in the work that we mentioned that presented the t-way coverage algorithms, their basic assumption was that the SUT was already represented in the model (such as LTS), but with no recommendation as to how to get this model.

Our suggestion in this proposal is a comprehensive methodology for the test suite in three stages. The first stage begins with defining the test process as BP, based on system requirements. The second stage creates a model by Model-Checking, in which basically all of the test cases are represented as LTS. The third stage is based on that model, creating a minimum set of test cases using the t-way method that covers all scenarios. In addition we developed new methods for t-way test case generation for minimal coverage by using solvers such as the 
`z3' library in Python. This method has two steps, such as presented in Kuhn’s work (Target and Testing Sequences), but unlike their work our focus is representing the system model in the solver, and doing so as a regular expression (RegEx). The first and second stages in our methodology harness and utilize the capabilities of BP, and the third stage is based on Yu et al.~\cite{Yu2012} in the Kuhn group works and his followers~\cite{Sheng2018,Duan2019, Bombarda2020}, while in our early work we suggested solving it using a 'z3' solver. Our suggested approach positions a number of advantages over what has been presented so far. By using BP infrastructure, we are able to model composite and large systems, e.g., on-board satellite software~\cite{Bar-Sinai2019}, and we use a common python library as a solver that was proven for that system. For the minimal coverage problem that we present, we say that $L'$ is a t-way coverage of $L$ if:
\begin{equation} \label{eu_eqn}
\forall \sigma_1 \cdots \sigma_t \in \Sigma^t  \quad 
(\Sigma^*\sigma_1\cdots\Sigma^*\sigma_t\Sigma^*)\cap L \neq \emptyset \quad \Rightarrow \quad (\Sigma^*\sigma_1\Sigma^*\cdots\Sigma^*\sigma_t\Sigma^*) \cap L' \neq  \emptyset
\end{equation}

\subsubsection{Test case coverage – our examples}
Because of the importance to the issue of the requirements coverage, we have examined a number of options for coverage. The following are two representative examples in our opinion as a basis for our research.

The first example that we implemented was ‘IBM ponder’ (January, 2013)~\cite{IBMResearch}. 

\begin{figure}[h]
\centering
\includegraphics[scale=0.7]{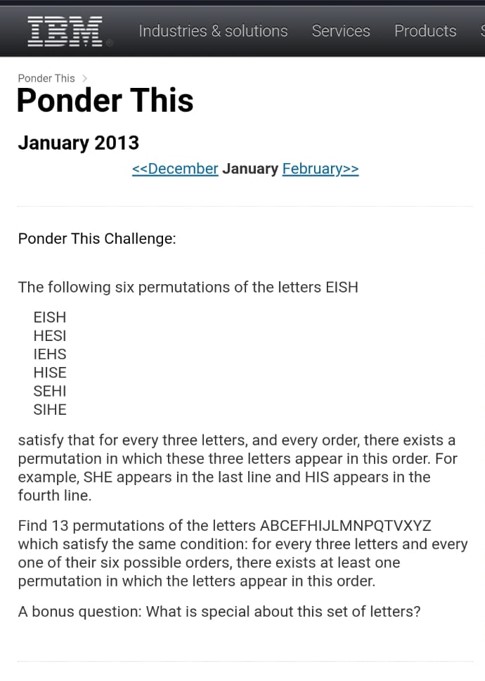}
\caption{IBM Ponder.}
\label{fig:ibm}
\end{figure}

We tried to crack a riddle that was published by Prof. Margalit on the IBM Israel website. The ponder (riddle) is shown in Figure~\ref{fig:ibm}. We found that this riddle represents our test case coverage problem and the same algorithms should solve both of them. The riddle challenged the manner of generating coverage of every three-letter word in any order by at least once in a list of letter combinations, and minimized the number of combinations. The analogy of the testing of this riddle is that each letter represents a possible parameter or possible state of the SUT and is required to produce a minimum list of test cases (each 'word' is a test case) that covers all of the generated 3-way words. In this example we didn't use BP as the first and second stages because the output was given (a long word with 18 different letters). We divided it into two parts; first a generated list of all t-way possibilities, and second to found the minimal number of words, and the permutation of the initial word, that covers all of the words listed in the first part. An example of solution source code is shown in Figure~\ref{fig:ibm}. In both parts we generated a RegEx that relaxed the required solution and ran the solver 'z3' to find them. We represent the riddle as a general notation:

\begin{align} \label{eq:2}
\left\{\Pi_{\Sigma'}(w)\colon w\in L'\right\} &= \left\{\Pi_{\Sigma'}(w)\colon w \in L\right\} \\
\Pi_{\Sigma'}(w) &= \begin{cases} 
    w[1] \Pi_{\Sigma'}(w[2..]), &  \text{if } w[1] \in \Sigma' \\ 
    \Pi_{ \Sigma'} (w[2..]),    &  \text{if } w[1]  \notin \Sigma'
\end{cases}\\
L  &= \Sigma! 
\end{align}

The RegEx for the first part (in 'z3' notation) is:

\begin{figure}[h]
\includegraphics[scale=0.65]{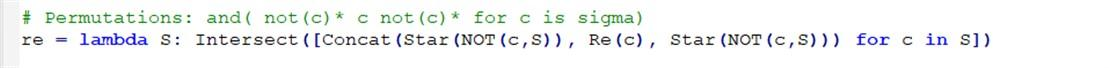}
\label{fig:perm1}
\end{figure}

NOT(c,S) is defined as:

\begin{figure}[h]
\includegraphics[scale=0.7]{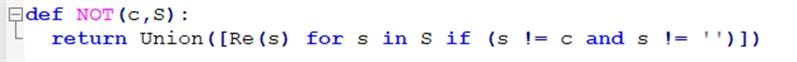}
\label{fig:perm2}
\end{figure}

And, the RegEX for the second part (in `z3' notation) is:
\begin{figure}[h]
\includegraphics[scale=0.65]{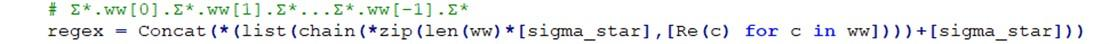}
\label{fig:perm3}
\end{figure}
Perhaps modelling the SUT as a RegEx presentation in characters (letters) is not the best example of how a solver such as `z3' can solve composite problems, but at this point in our research it is good enough. 

Source code demonstration is shown in Appendix C.

\begin{figure}
\centering
\begin{subfigure}{.5\textwidth}
  \centering
\includegraphics[scale=0.45]{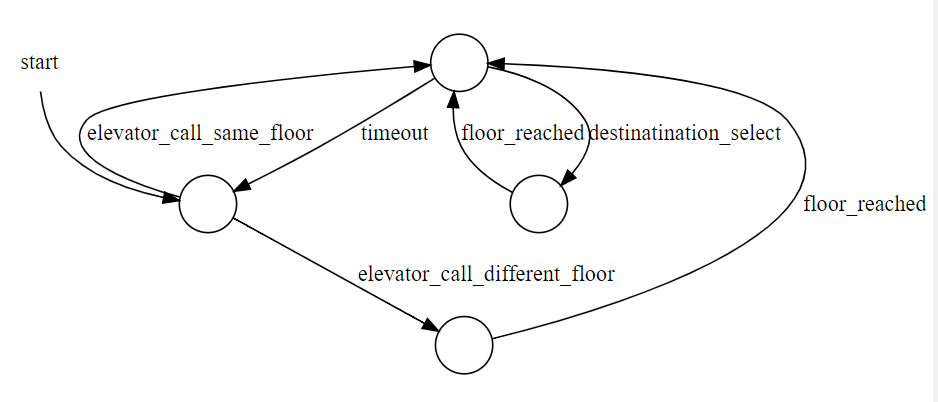}
\end{subfigure}%
\begin{subfigure}{.5\textwidth}
  \centering
\includegraphics[scale=0.4]{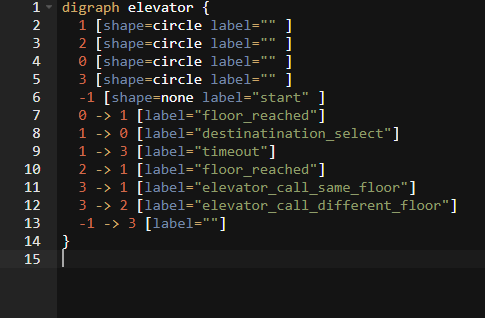}
\end{subfigure}
\caption{Elevator automaton graph.}
\label{fig:auto}
\end{figure}

\begin{figure}[h]
\centering
\includegraphics[scale=0.5]{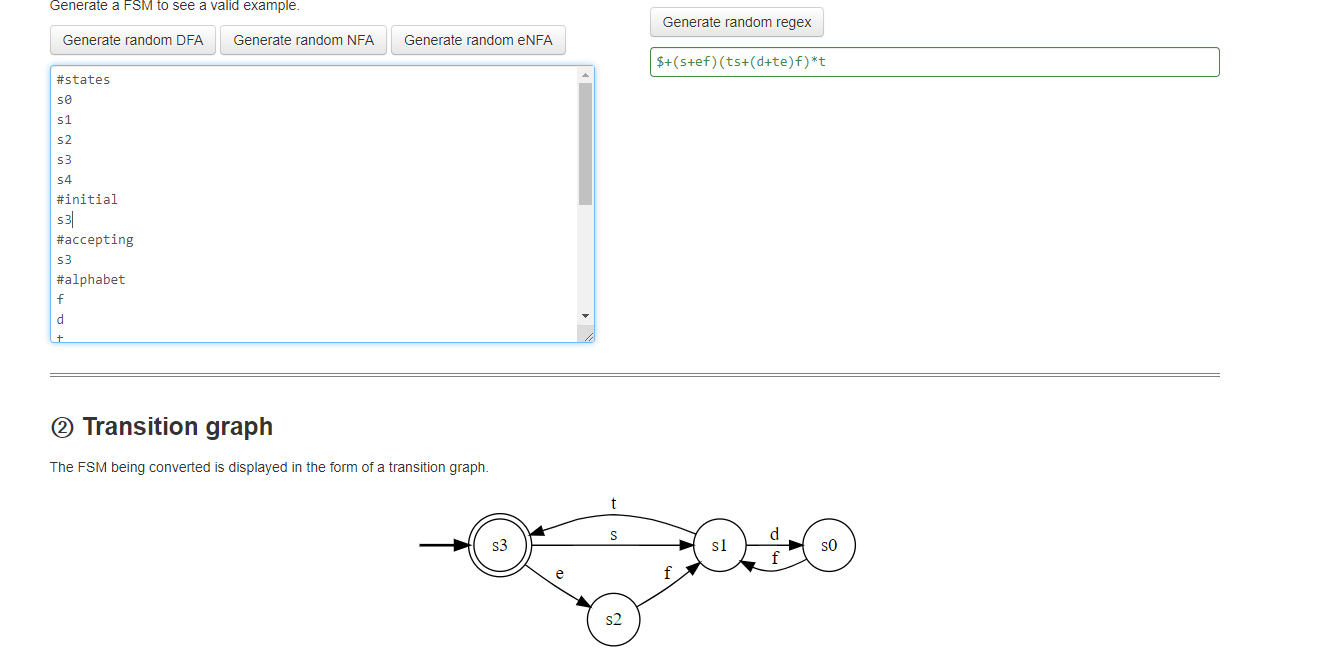}
\caption{Elevator translation to RegEx.}
\label{fig:regex}
\end{figure}

In the second example, we try to mimic the coverage process based on Kuhn's followers~\cite{Duan2019}. We tried to mimic one of the latest works based on Kuhn's followers who represent the SUT with automata. In our work we demonstrate the full process in two simple examples that they presented: vault and elevator. We implemented as proof of concept (POC) our proposed research in a three-stage process. In the first stage we implement these two examples in BP and generate b-threads, b-sync, Requests, Waitfor and Block, which mimic the automata like they did; In the second stage we run a model-check in a very specific way that generates an automaton (in graph format) from the BP program. The output is a finite state machine (FSM) representing automaton. Using graphvizOnline website~\cite{Graphviz} we present the automaton in Figure~\ref{fig:auto}. From this graph we copy the automaton and convert the automaton format as required, using fsm2regex website~\cite{FSM2Regex}, and generate RegEx, as shown in Figure~\ref{fig:regex}. In the third stage we convert the format of the regex generated by fms2regex to `z3' notation and run the Z3 solver to find the minimum test cases that cover the examples.
 
\subsection{POC – feasibilities studies}
\subsubsection{Telephony system – online concept, closed-loop}
\begin{figure}[h]
\centering
\includegraphics[scale=0.4]{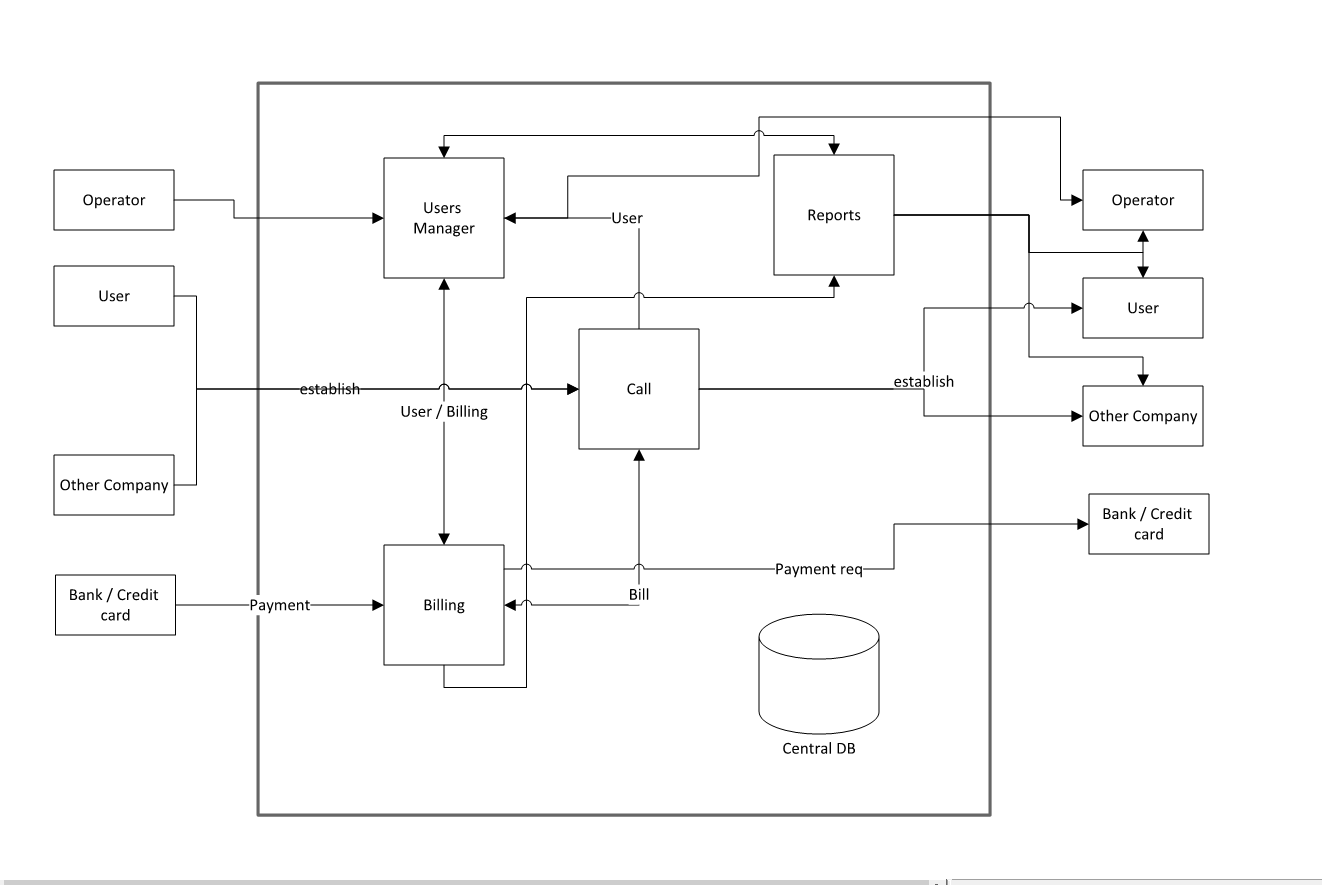}
\caption{Telephony system - system diagram.}
\label{fig:TS}
\end{figure}

We started by developing a simulation of a Telephony system (TS) that was built as a first playground that we used to adopt the BP principles in the system testing arena. The telephony system contains the following entities and capabilities: Telephony company Users - add, update and delete users; Establish a call between users of different call types (domestic, international, collect or free); Send an SMS between users in different types (SMS or MMS); Users Bill – change and check; internal Telephony database that contained telephony system operational data such as users, calls, SMSs, and tariffs. Figure ~\ref{fig:TS} shows the telephony system.

The first requirement that we define which we like to test is: "After each phone call that is established, the paid user's charge per call in the correct amount is billed based on the given rate." 

Aside from the telephony system, we developed the automated testing tools. The testing tools used BPjs framework. The testing tools simulate the telephony systems' users by applying the TS APIs. Under BPjs we simulated the users adding, between users we randomly selected, established calls (the call parameter, call type and call length were chosen randomly). The testing system tests, periodically (by 'testBill' request), the bill that each user has charged. 
Each one of these capabilities was a b-thread, and each activity reacted in a b-sync mechanism. In TS we try to generate the test cases as reacting to the TS action: the user's bill was changed (in the testing system) whenever a call was established, as defined in the requirement. The method for testing was to periodically test the user's bill. The TS compares the referenced calculated bill in the testing system to the user's bill in telephony DB. However, the TS skips testing when there is a known possibility that the bill is wrong. Then we divided the action into two separate threads: the first was to periodically test the bill and the second was to establish calls and update the user's bill accordingly.

The first thread is:  "\underline{Correct amount is billed?}":

{\em \color{blue}

\hspace{1cm}int amt = 0;

\hspace{1cm}\color{red}wait: \color{blue}creactUser(u1?)

\hspace{1cm}forever

\hspace{1.5cm}try:
   
\hspace{2cm}\color{red}request: \color{blue}testBill(u1, amt)
	
\hspace{1.5cm}\color{red}break upon: \color{blue}

\hspace{2cm}updateBill((u1, y?);
	
\hspace{2cm}amt += y
	                       
}

The second thread is: "\underline{Charge Per {\bf call}}"
{\em \color{blue}

\hspace{1cm}forever:

\hspace{2cm}\color{red}wait: \color{blue}call (u1, u2?);
	
\hspace{2cm}\color{red}request: \color{blue}updateBill(u1; f(u1, u2));
}

The second case that we checked affected the tests by adding requirements such as "After each SMS sent between two users, the paid user's bill \underline{charge per SMS} in the \underline{correct amount is billed} based on the given rate." As expected, this new requirement doesn't affect the existing test in either thread; we therefore added the 3rd thread to the testing system.

The third thread is: "\underline{Charge Per {\bf SMS}}"
{\em \color{blue}

\hspace{1cm}forever:

\hspace{2cm}\color{red}wait: \color{blue}sms (u1, u2?);
	
\hspace{2cm}\color{red}request: \color{blue}updateBill(u1; f1(u1, u2));
	
}
The method that we try to elaborate in the telephony system is that the test thread reacts based on the action in the SUT. We called this method the 'online' testing method or 'closed-loop' testing. 

The advantage of this method is that the basic test threads (i.e. "charge per call", "charge per sms" or "correct amount is billed?") are generated and they act upon action from the SUT. That means that test scenarios are generated 'on the fly' and try to cover the scenarios or the use-cases in the SUT in a way that it is done in the "real world". 

The major disadvantage is that there is no way to model checking on test scenarios. The graph, built from the b-thread tests, let us trace options on the test scenarios and find the best scenarios to run accordingly. In the 'on-the-fly' method, where the test cases and the b-threads are built while the test is done, it is impossible.

\subsubsection{Cucumber / Gherkin} 
Another path that we tried to research was the way to use Off the Shelf (OTS) tools as DSL for foreground (or frontend) and to use Gherkin as an actuator tool. We check Cucumber and Gherkin as candidates.

Cucumber is a testing tool that supports the Behaviour Driven Development (BDD) framework. It defines application behaviour using simple English text, defined by a language called Gherkin. The way that Cucumber works is that Cucumber reads the code written in plain English text (i.e. in Gherkin) in the 'feature' file. It finds the exact match of each step in the step definition (a source code file). The piece of code to be executed can be different software frameworks like JavaScript, Selenium, etc. Each feature file is one 'Feature' and may contain one or more 'Scenarios'. Each scenario is built from 'Steps'. Each scenario is a test scenario that has the test logic. The vocabulary for the logic step is the following idioms: Background, Give, When, Then ('And' is the same as the previous one). Gherkin is able to use words in the steps as variables. These variables come between double quotation marks ("xxx" or "123").

The following is an example of a scenario written in Gherkin as part of Telephony system test: 

{\em

\hspace{1cm}\color{red}Scenario: \color{blue}charge Per Call

\hspace{2cm}\color{red}Then \color{blue}forever
	
\hspace{2cm}\color{red}Given \color{blue}"call" in "Calls"

\hspace{2cm}\color{red}Then \color{blue}calculate charge

\hspace{2cm}\color{red}And \color{blue}update bill according "call" type
}

For each scenario and for each step line in a scenario, the Cucumber finds the function that fits, for example, for the step 'Given "call" in "Calls" ':
{\em

\hspace{1cm}\color{red}@Given \color{blue}("{string} in {string}")
   
\hspace{1cm} public void givenObjInClass(String obj, String inClass)

\hspace{1cm} \{	
   
\hspace{2cm}if (obj.equals("usr") \&\& inClass.equals("User"))

\hspace{2cm}\{

\hspace{3cm}//Do something when "usr" in "User"

\hspace{2cm}\}
	
\hspace{2cm}else if (obj.equals("call") \&\& inClass.equals("Calls"))
	
\hspace{2cm}\{
	
\hspace{3cm}//Do something when "call" in "Calls"
        
\hspace{2cm}\}
	
\hspace{1cm}\}

}

Or for step ' Then calculate charge'
{\em

\hspace{1cm}\color{red}@Then \color{blue}("calculate charge")

\hspace{1cm}public void chargeBillCalc()
   
\hspace{1cm}\{
   
\hspace{2cm}//Do something 

\hspace{1cm}\}
}

We try to generate matching mapping between Gherkin Idioms and BP idioms such as:
Scenario generates b-thread; 'Given' generates b-sync with 'Request'; 'When' generates b-sync with 'Waitfor'; and 'Then' is the test's logic. But this mapping was poor for all vocabulary in BP, especially the more powerful BP idioms such as 'Block', 'Break-upon' and 'Interrupt'.

We found out that the power of Cucumber and Gherkin, that we are going to adopt in our methodology, will be the separation of the testing into (at least) two levels of testing implementation. The first level is the scenario process level (like feature file level) and the second level is the handling level (like the code file level).

\subsection{Proof-of-concept (POC) tools}
The POC implementation demonstrates all of the above mentioned contributions.

\begin{figure}[h]
\centering
\includegraphics[scale=0.6]{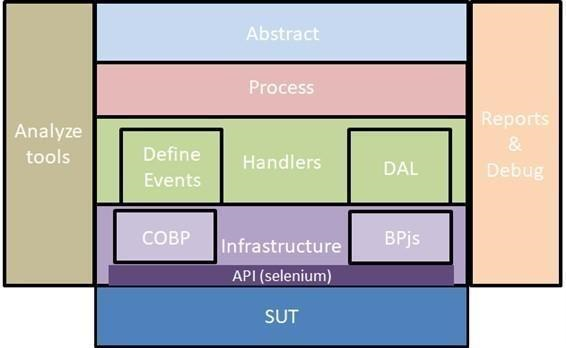}
\caption{POC framework structure.}
\label{fig:POC}
\end{figure}

In order to prove the feasibility of the method of using BP to design system tests and produce a graph of all the possible states from which the test scenarios are constructed, we took one use case and applied the method in a limited way. We were able to realize and base the research on this success. Our proposal in this work is to build a tool kit on four levels. Each level serves a different purpose and a different role, and enables communication between each two neighbours on the different levels. Figure~\ref{fig:POC} showed the POC structure of its components, the layers on which it rests and supporting tools. This proposal is based on the proof of concept (POC) demo that we developed.
\begin{enumerate}
\item Infrastructure level - This is the lower level. This level serves all other levels and has a common infrastructure for all kinds of automated testing using this tool kit and will be adjusted by the software engineers or automated test developers. The infrastructure contains all functions that actually activate the test process (e.g., runs a web browser by selenium API). This level is mostly implemented based on BPjs and COBP and uses the context mechanism b-thread, b-sync, with all BP vocabulary.
\item Handler level - This is the second level. On this level we define the events (in POC these functions are called ``define event'') run at each test step in the test case and include the test’s internal logic, such as the order of filling in fields on a web form (the process level is on the 3rd level). The development of this level is executed by an automated test developer. Each event definition group of functions handles one object (e.g., form process) and is very specific to the test case and to the system under test (SUT). 
\item Process level – This is the third level. This level contains the business-logic in the test case process and selection of which test case will be executed. The process level mimics the 'feature' definition in the Cucumber / Gherkin infrastructure. The process level defines the processes that require testing and the test cases and the process logic.  We could generate end-to-end tests or just one process test; a sanity test or negative test. This should be the level that the system engineers and the tests engineers can write or can discuss and which documents the traceability between the system’s requirements and specifications and system test cases. At this level the test definitions should include all of the capabilities derived from BP methods. Using Request, WaitFor, Block, Break-Upon and Interrupt, using context to manage the system state and test data along the test process. At this level the model check could analyse the test processes.  
\item Abstraction level – This is the fourth level. This is the level that defines the test on the same one as the process level. The tests that are defined here will be translated automatically to the tests process. In the abstraction level, the tests will be written using a new technique such as diagrams or formal language defined by DSL and will not mention a programming language. The DSL will have logical structures and repetitions but without the form of "if…then…else" or "for/while/do loops". 
\end{enumerate}
Appendix C shows details and source code capturing the POC implementation. We developed one test case on the Moodle website as preliminary study for this proposal.
\printbibliography %

\end{document}